\documentclass[a4paper]{spie}  

\usepackage[]{graphicx}

\usepackage{fleqn,epsfig,color,graphicx,url,psfrag}
\usepackage{amsfonts,amssymb,amsmath,esint}

\newcommand{\bil}[2]{a\left(#1,#2;\nu\right)}

\newcommand{\fprim}[1]{f\left(#1\right)}

\newcommand{\fdual}[2]{l^{\mathrm{o}}_{#1}\left(#2\right)}

\newcommand{\myspan}[1]{\mathrm{span}\left\{#1\right\}}
\newcommand{\myset}[1]{\left\{#1\right\}}

\newcommand{\nrb}{N}
\newcommand{\nn}{\mathcal{N}}

\newcommand{\uex}{u}
\newcommand{\un}{u^\nn}
\newcommand{\urb}{u^\nrb}

\newcommand{\sn}[1]{s^\nn_{#1}(\nu)}
\newcommand{\srb}[1]{s^\nrb_{#1}(\nu)}

\newcommand{\mani}{\mathcal{M}^\nn}

\newcommand{\xn}{X^\nn}
\newcommand{\xrb}{X^\nrb}

\newcommand{\xdualn}{\left(X^\nn\right)'}

\newcommand{\lrb}[1]{W^\nn_{#1}}

\newcommand{\rprimrb}[1]{r^{\mathrm{pr}}\left(#1;\urb;\nu\right)}

\newcommand{\norm}[2]{\left|\left|#1\right|\right|_{#2}}
\newcommand{\real}{\mathbb{R}}
\newcommand{\complex}{\mathbb{C}}

\newcommand{\curl}{\mathbf{curl}\;}

\newcommand{\hcurl}{\mathrm{H}\left(\curl,\Omega\right)}

\newcommand{\xitrain}{\Xi_{\mathrm{train}}}

\def\myspace{\vspace{0.5cm}\noindent\\}

\title{Reduced basis method for computational lithography}

\author{
Jan Pomplun\supit{\,ab},
Lin Zschiedrich\supit{\,ab},
Sven Burger\supit{\,ab},
Frank Schmidt\supit{\,ab}
\skiplinehalf
\supit{a}
Zuse Institute Berlin,
Takustra{\ss}e 7,
D\,--\,14\,195 Berlin,
Germany
\smallskip\\
\supit{b}
JCMwave GmbH,
Haarer Stra{\ss}e 14a,
D\,--\,85\,640 Putzbrunn, 
Germany
}

\authorinfo{
Corresponding author: J. Pomplun\\
URL: http://www.zib.de/nano-optics/\\
Email: pomplun@zib.de
}

\begin{document} 
  \maketitle 


\noindent
This paper has been published in Proc.~SPIE Vol. {\bf 7488}
(2009) 74882B, 
({\it Photomask Technology 2009, Larry S. Zurbrick; M. Warren Montgomery, Editors})
and is made available as an electronic preprint with permission of SPIE. 
Copyright 2009 Society of Photo-Optical Instrumentation Engineers. 
One print or electronic copy may be made for personal use only. 
Systematic reproduction and distribution, duplication of any material in this paper for a fee or for 
commercial purposes, or modification of the content of the paper are prohibited. 

\begin{abstract}
A bottleneck for computational lithography and optical metrology are long computational times for near field simulations. For design, optimization, and inverse scatterometry usually the same basic layout has to be simulated multiple times for different values of geometrical parameters.

The reduced basis method allows to split up the solution process of a parameterized model into an expensive offline and a cheap online part. After constructing the reduced basis offline, the reduced model can be solved online very fast in the order of seconds or below. Error estimators assure the reliability of the reduced basis solution and are used for self adaptive construction of the reduced system.

We explain the idea of reduced basis and use the finite element solver JCMsuite constructing the reduced basis system. We present a 3D optimization application from optical proximity correction (OPC).
\end{abstract}


\section{Introduction}
The importance of numerical simulations for design, optimization and metrology of photomasks has grown rapidly over the last years due to the ongoing miniaturization of integrated curcuits.

A bottleneck for these many-query and real-time applications are long computational times for rigorous simulations of the near field in the photomask. Usually the same basic layout has to be simulated multiple times for different values of geometrical parameters, e.g. line width, absorber edge angle, etc.

The reduced basis method \cite{Pat07a,ROZ08,ZHE08} can be applied to this task. The solution process is decomposed into an expensive offline and a cheap online step. In the offline step the reduced basis is built self-adaptively by solving the underlying model rigorously several times. The full model is then projected onto the reduced basis. In the online step the assembled reduced system can be solved in the order of seconds independent on the size of the original problem. Furthermore methods from the well established field of a posteriori error estimation of finite element methods \cite{AIN00} can be applied to the reduced basis method to assure the reliability of the computed output and also for construction of the reduced basis \cite{Pomplun2008a,CHE08}. These are advantages in comparison to interpolation or table based methods for output data.

\begin{centering}
\begin{figure}
(a)\hspace{10cm}(b)\vspace{1.0cm}\\
\psfrag{dx}{$d_{x}$}
\psfrag{dy}{$d_{y}$}
\psfrag{p1y}{$p_{1,y}$}
\psfrag{p2y}{$p_{2,y}$}
\psfrag{p1x}{$p_{1,x}$}
\psfrag{p2x}{$p_{2,x}$}
\includegraphics[height=6cm]{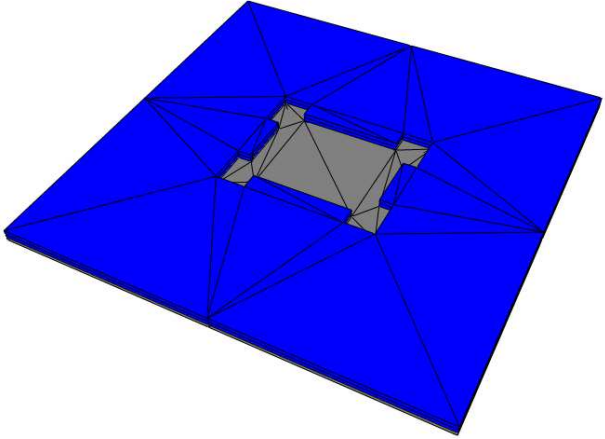}\hfill
\includegraphics[height=6cm]{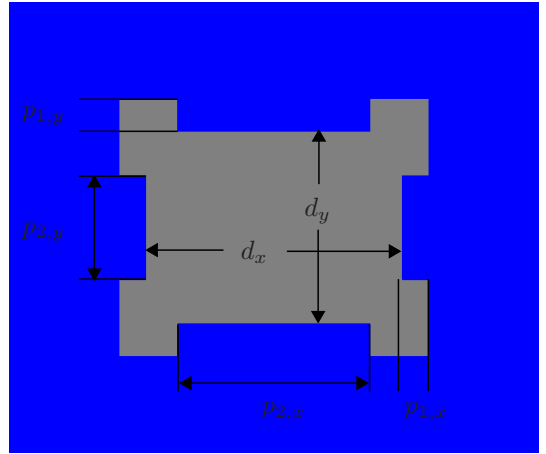}
\caption{\label{fig:opcMaskModel}
(a) 3D mask model used for finite element computation. (b) Definition of OPC parameters used for optimization.
}
\end{figure}
\end{centering}

In the following we first describe the basics of the reduced basis method and then apply it to an OPC optimization problem of a contact hole array depicted in Fig. \ref{fig:opcMaskModel}.

\section{Input-output relationship}
\label{sec:ior}
Before explaining the reduced basis method we first derive the mathematical setup for the scattering problem. Suppose we have a photomask with a parameterized geometry as depicted in Fig. \ref{fig:opcMaskModel}. In the following the free geometrical parameters are denoted by a $p$-tuple $\nu$ in a bounded parameter space $D\subset\real^{p}$. Fixing the input parameters $\nu$ to the mask model, the scattering problem can be solved numerically with a Maxwell solver relying, e.g. on the finite element method (FEM), and outputs of interest, e.g. diffraction orders, can be computed. Hence the problem can be formulated as an input output relationship. In this example the input is a set of geometrical parameters, the output are the diffraction orders. For notational convenience we consider a single output of interest in the following. Mathematically the input-output relationship is given as follows:\\\noindent
Given geometrical parameters $\nu\in D$ determine the output of interest $\sn{}$:
\begin{align}
  \label{eq:ooin}
  \sn{}=\fdual{}{\un(\nu)},
\end{align}
where $\un(\nu)$ is the solution to following problem:\\\noindent
Find $\un\in \xn$ such that:
\begin{align}
  \label{eq:n}
  \bil{\un}{v}=\fprim{v}\,,\quad\forall v\in \xn.
\end{align}
\myspace
The electromagnetic scattering problem \eqref{eq:n} is stated in so called ``weak form'' \cite{MON03} . Roughly spoken the sesquilinear form $\bil{\cdot}{\cdot}$ presents Maxwell's equations and the linear functional $\fprim{\cdot}$ the incoming field, e.g. a plane wave. The finite element space $\xn\subset\hcurl$ is the space in which the electric field solution $\un(\nu)$ is determined, where $\Omega$ is the domain of interest. Since Maxwell's equations and therewith the sesquilinear form are parameter dependent, the solution $\un$ is also parameter dependent. The output of interest, e.g. a diffraction mode in a periodic setting, is described via a linear output functional applied to the electric field $\un$:
\begin{align}
\begin{split}
  l^{o} : \xn\rightarrow& \complex,\\
\un(\nu)\mapsto& \fdual{}{\un(\nu)}=\sn{}.
\end{split}
  \label{eq:ooiFunc}
\end{align}
Hence $l^{o}\in\xdualn$, which is the dual space of $\xn$ \cite{REE80}. The discretized version of Maxwell's equations \eqref{eq:n} will be referred to as ``truth approximation''. Usually the truth approximation has to be solved several times for different parameter values $\nu$ in design and optimization applications or inverse scatterometry. Typically $\nn$ is very large such that already a single solution process takes very long which prohibits many-query or real-time application.
\section{Reduced basis method}
\label{sec:rbm}
The purpose of the reduced basis method is to construct an approximative input-output relationship which can be solved very fast. Furthermore it is important to construct error estimators which control and quantify the quality and reliability of the reduced basis solution. Finally we want that computation of the reduced basis solution and error estimation becomes independent on the number of finite element degrees of freedom $\nn$.

The idea of the reduced basis method is simple \cite{POM09b}. Let us denote the manifold of all possible solutions of the truth approximation \eqref{eq:n} by:
\begin{align}
  \label{eq:defMani}
\mani=\myset{\un(\nu) \mbox{ is a solution to \eqref{eq:n}}\;|\;\nu\in D}.
\end{align}
Now suppose $\mani$ can be approximated by a low dimensional space $\xrb\approx\mani$, with
\begin{align}
  \label{eq:dimrb}
  \dim \xrb=\nrb\ll\nn.
\end{align}
Then it is reasonable to assume that the following reduced basis approximation to the truth approximation will give results of good quality:\\\noindent
Given geometrical parameters $\nu\in D$ determine the output of interest $\srb{}$:
\begin{align}
  \label{eq:ooirb}
  \srb{}=\fdual{}{\urb(\nu)},
\end{align}
where $\urb(\nu)$ is the solution to following problem:\\\noindent
Find $\urb\in \xrb$ such that:
\begin{align}
  \label{eq:rb}
  \bil{\urb}{v}=\fprim{v}\,,\quad\forall v\in \xrb.
\end{align}
\myspace
The space $\xrb$ is the reduced basis space and is a subspace of the finite element space $\xn$.

\subsection{Reduced basis space}
An important question which arises is how to construct the reduced basis space $\xrb$ such that it gives a good approximation to the space of all possible solutions $\mani$. To answer this question we first define a sequence of hierarchical subsets of the parameter domain $D$. Let $\nu_{i}\in D\,,\;i=1,\dots,\nrb_{\mathrm{max}}$. Then we define:
\begin{align}
  \label{eq:defNestPar}
S_{i}=&\myset{\nu_{1},\dots,\nu_{i}}\,,\quad i=1,\dots,\nrb_{\mathrm{max}}.
\end{align}
These sets have the following property:
\begin{align}
  \label{eq:nestParHier}
S_{1}\subset S_{2}\subset\dots\subset S_{\nrb}\subset\dots\subset S_{\nrb_{\mathrm{max}}}.
\end{align}
The Lagrange reduced basis space $\lrb{\nrb}$ of dimension $\nrb$ is then defined by:
\begin{align}
  \label{eq:defLagSpace}
\lrb{\nrb}=&\,\myspan{\un(\nu) \mbox{ is a solution to \eqref{eq:n}}\;|\;\nu\in S_{\nrb}},
\end{align}
hence it is spanned by solutions to the truth approximation for fixed parameter values. These solutions are called snapshot solutions. It is also possible to include first and higher derivatives of the field $\un$ with respect to parameters into the reduced basis space, which leads to so called Taylor and Hermite spaces. Here we notice the expensive ``offline'' costs of the reduced basis method. For construction of a Lagrange reduced basis space of dimension $\nrb$ the truth approximation, i.e. the full problem, has to be solved $\nrb$ times. However the reduced basis space has only to be assembled once and in principle a parallelization of this process is possible since computation of different snapshots is independent of each other.
\subsection{Reduced basis system}
In the following we construct the explicit form of the reduced basis system \eqref{eq:rb} which has to be solved in the online phase. Suppose we have given a basis 
\begin{align}
  \label{eq:rbrb}
  B^{\nn}_{\nrb}=&\myset{\zeta_{q}^{\nn}|\,q=1,\dots,\nrb}
\end{align}
of $\xrb$. Then we can expand the reduced basis solution into this basis:
\begin{align}
  \label{eq:rbexp}
  \urb(\nu)=&\sum\limits_{q=1}^{\nrb}\alpha_{q}(\nu)\zeta_{q}^{\nn}.
\end{align}
Online the parameter dependent coefficients $\alpha_{q}(\nu)$ have to be computed. Therefore we insert ansatz \eqref{eq:rbexp} into the reduced basis system \eqref{eq:rb} and use the fact that if \eqref{eq:rb} holds for all $v\in B^{\nn}_{\nrb}$ then it holds for all $v\in\xrb$:
\begin{align}
  \label{eq:rbEqn}
\sum\limits_{q=1}^{\nrb}\alpha_{q}(\nu)  \bil{\zeta_{q}^{\nn}}{\zeta_{n}^{\nn}}=\fprim{\zeta_{n}^{\nn}}\,,\quad n=1,\dots,\nrb,
\end{align}
which gives a linear system of equations for the coefficients $\alpha_{q}(\nu)$. System \eqref{eq:rbEqn} has to be assembled and solved online. On the left hand side we have a parameter dependent matrix:
\begin{align}
  \label{eq:rbMat}
A^{\nrb}(\nu)=&\left(\bil{\zeta_{q}^{\nn}}{\zeta_{n}^{\nn}}  \right)_{q,n=1,\dots,\nrb}\;,
\end{align}
and on the right hand side a parameter independent vector:
\begin{align}
  \label{eq:rbVec}
f^{\nrb}=&\left(\fprim{\zeta_{n}^{\nn}} \right)_{n=1,\dots,\nrb}\;.
\end{align}
The vector $f^{\nrb}$ can be assembled offline. However computation of each matrix entry:
\begin{align*}
  \bil{\zeta_{q}^{\nn}}{\zeta_{n}^{\nn}}  
\end{align*}
depends on the dimension of $\zeta_{q}^{\nn}$, which is the number of finite element degrees of freedom $\nn$. Since this assembling step has to be performed online we want to avoid any $\nn$ dependence. In the following we explain how we can perform an online-offline decomposition of the assembling step.
\subsection{Online-offline decomposition}
For online-offline decomposition of the assembling step \eqref{eq:rbMat} we need an affine decomposition of the system sesquilinear form $\bil{\cdot}{\cdot}$, which is defined as follows:
\begin{align}
  \label{eq:affBilin}
\bil{v}{\uex}=\sum\limits_{m=1}^{Q}\Theta_{m}(\nu)a_{m}(v,\uex),
\end{align}
where $\Theta_{m}(\nu)$ are parameter dependent functions and $a_{m}(\cdot,\cdot)$ parameter independent sesquilinear forms. If we can construct such a decomposition the following parameter independent matrices can be assembled offline:
\begin{align}
\nonumber
  A^{\nrb}_{m}=&\left( a_{m}(\zeta_{q}^{\nn},\zeta_{n}^{\nn})\right)_{q,n=1,\dots,\nrb}\,,\quad m=1,\dots,Q.
\end{align}
The parameter dependent system matrix \eqref{eq:rbMat} is then assembled online according to:
\begin{align}
\nonumber
  A^{\nrb}(\nu)=&\sum\limits_{m=1}^{Q}\Theta_{m}(\nu)A^{\nrb}_{m}.
\end{align}
The costs are $O(\nrb^{2}Q)$, where $\nrb$ is the reduced basis dimension and $Q$ the number of terms in the affine decomposition \eqref{eq:affBilin}. The solution of the reduced basis system has costs $O(\nrb^{3})$. For computation of the output of interest we get the following decomposition:
\begin{align}
  \srb{}=&\fdual{}{\urb(\nu)}\nonumber\\
=&\fdual{}{\sum\limits_{q=1}^{\nrb}\alpha_{q}(\nu)\zeta_{q}^{\nn}}\nonumber\\
=&\sum\limits_{q=1}^{\nrb}\alpha_{q}(\nu)\fdual{}{\zeta_{q}^{\nn}},
  \label{eq:ooiDecomp}
\end{align}
where the quantities $\fdual{}{\zeta_{q}^{\nn}}$ can also be computed offline. For the output of interest we have costs $O(\nrb)$. Hence with an affine decomposition \eqref{eq:affBilin} the costs of solving the reduced basis system are independent on $\nn$ as desired.

In \cite{Pomplun2008a} we show how an affine decomposition can be constructed for electromagnetic scattering problems.
\subsection{Error estimation}
Since the reduced basis method is an approximation technique it is important to be able to quantify the reliability of the output. Let us define the error of the reduced basis solution by:
\begin{align}
  \label{eq:error}
  e=\un-\urb.
\end{align}
Then it can be shown \cite{MAD01,ROZ08} that this error is bounded by:
\begin{align}
\norm{e(\nu)}{\xn}&\le\Delta(\nu),\nonumber
\end{align}
with:
\begin{align}
\label{eq:errBound}
\Delta(\nu)&=\frac{1}{\beta(\nu)}\norm{\rprimrb{\cdot}}{\xdualn}.
\end{align}
Here $\rprimrb{\cdot}$ is the so called primal residuum of the reduced basis solution $\urb$. The parameter dependent constant $\beta(\nu)$ is the so called inf-sup constant of the sesquilinear form $\bil{\cdot}{\cdot}$. It is important to note that all quantities of estimate \eqref{eq:errBound} can be computed online \cite{HUY07,ROZ08} without knowing the true solution. An appropriate online-offline decomposition furthermore makes computation of the bound independent on the number of finite element degrees of freedom such that an error bound can be computed for each reduced basis output online. Also this error estimator can be used in the offline construction of the reduced basis to explore the parameter space and find regions, where the reduced basis approximation has to be improved further, which is explained in the following section. This is a main advantage in comparison to interpolation methods. In \cite{POM09a} we will explain the subject of error estimation in more detail for electromagnetic scattering problems.

\begin{centering}
\begin{figure}
\begin{flushleft}
(a)\hspace{5.2cm}(b)\hspace{5.5cm}(c)\\\vspace{0.4cm}
\end{flushleft}
\centering
\psfrag{}{$$}
\includegraphics[height=4cm]{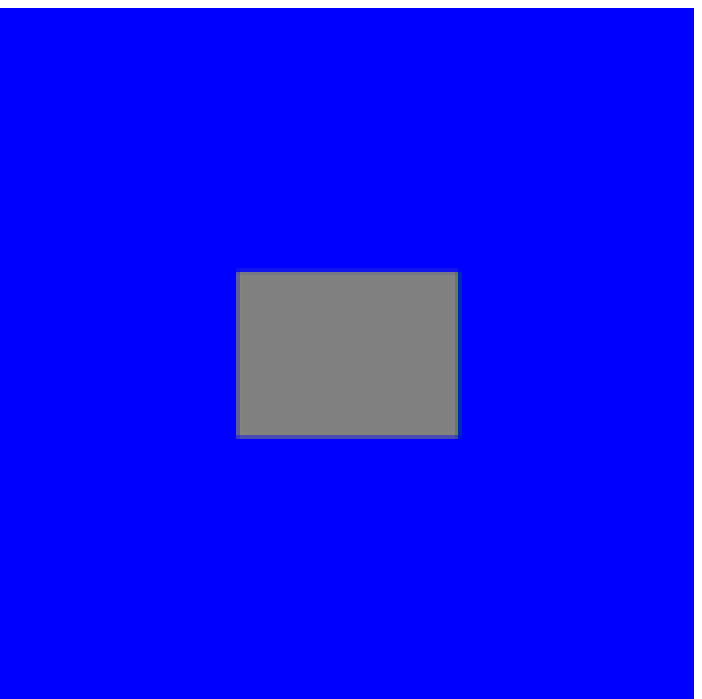}\hfill
\includegraphics[height=4cm]{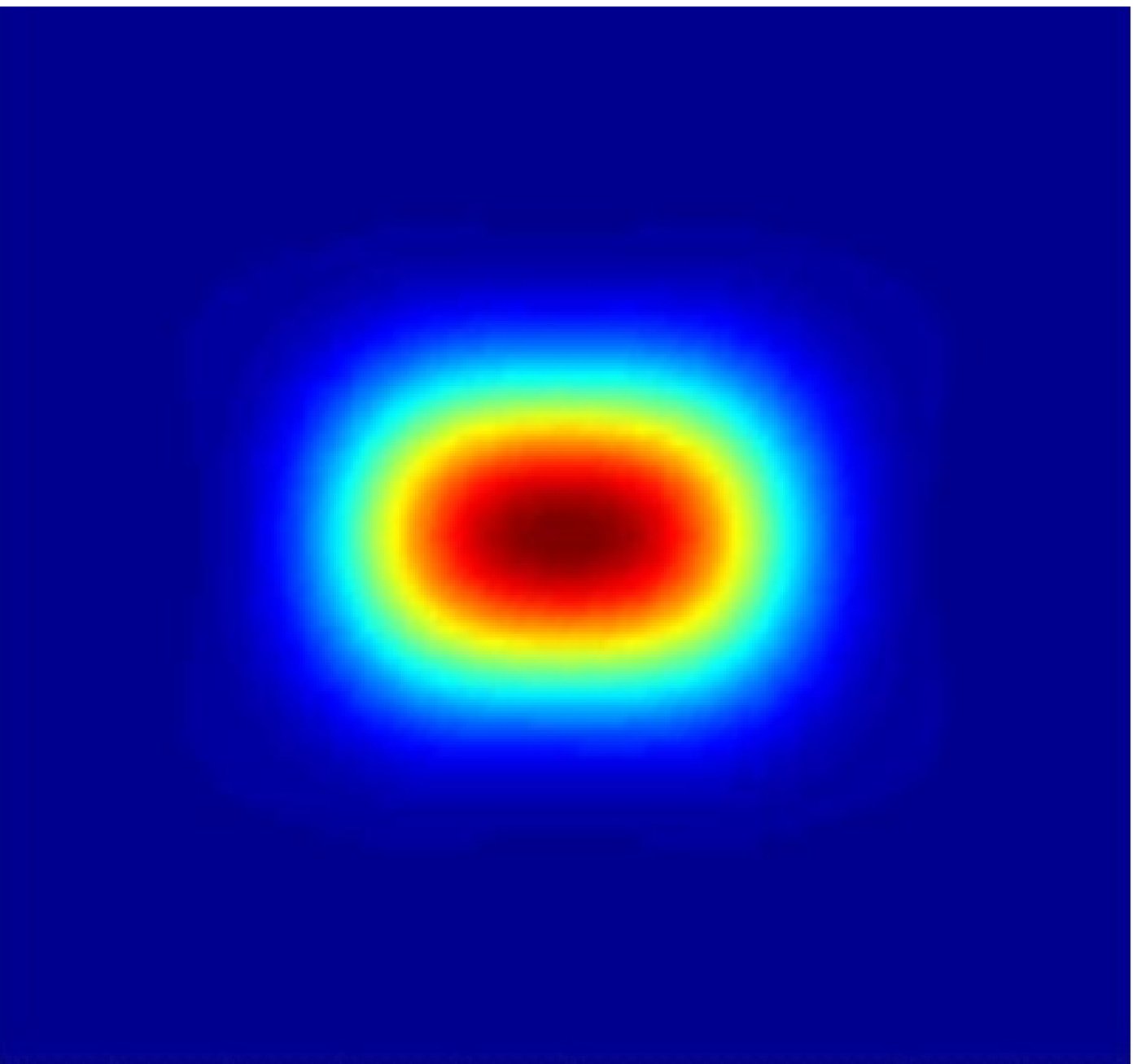}\hfill
\includegraphics[height=4cm]{fig/sourcesKSpace.eps}\hfill
\caption{\label{fig:opcMaskRound}
(a) Top view of mask layout for imaging of contact hole. Absorber material is shown in blue. (b) Far field image of contact hole shows corner rounding. (c) Position of incoming $k$-vectors used for conventional illumination.
}
\end{figure}
\end{centering}
\section{Greedy construction of reduced basis spaces}
\label{sec:greed}
Before we apply the reduced basis method we comment on the construction of the Lagrange reduced basis space $\xn=\lrb{\nrb}$ \eqref{eq:defLagSpace}. How to choose the snapshot parameters $\nu_{i}$ in \eqref{eq:defNestPar}? We use a greedy algorithm \cite{ROZ08} whose idea is given in the following. 

First we define a training set $\xitrain\in D$ of possible snapshot candidates. From this space we want to chose a number $\nrb$ of snapshot parameters for construction of the reduced basis. The first parameter $\nu_{1}$ is chosen randomly. Then we construct a one dimensional reduced basis approximation corresponding to the snapshot $\urb(\nu_{1})$. Now we evaluate the error estimator for this one dimensional reduced basis approximation on all candidate snapshots in the training sample $\xitrain$ and include the parameter value with the maximum error into the reduced basis because this is supposed to add a maximum of ''new information'' into the reduced basis. Then we have a two dimensional reduced basis and the process is continued iteratively. The process can be stopped e.g. if a certain maximum dimension is reached or if the error estimator gives sufficiently small bounds over the training set.
\section{Numerical examples}
Figure \ref{fig:opcMaskRound} shows the aerial image of a rectangular contact hole. In the far field the sharp corners are washed out which is refered to as corner rounding. Correction of this rounding can be achieved with OPC methods like the introduction of serifs to the corners of the contact holes as depicted in Fig. \ref{fig:opcMaskModel}. Our numerical example will be the optimization of this OPC structure to obtain a structure on the wafer which is closest to the desired rectangular shape.

The geometry of the contact hole is depicted in Fig. \ref{fig:opcMaskModel}. The size of the computational domain is $2.5\,\mu\mathrm{m}\times 2.5\,\mu\mathrm{m}$ with a total height of $90\,\mathrm{nm}$ including silica substrate and air. The height of the Chromium absorber is $50\,\mathrm{nm}$ with a refractive index of:
\begin{align*}
  n_{\mathrm{Cr}}=0.84-1.65i
\end{align*}
for a wavelength of $\lambda=193\,\mathrm{nm}$ \cite{LEV04}.
In $x$- and $y$-direction we apply periodic boundary conditions. The shape of serifs which we want to optimize is described by 4 input parameters $p_{1,x}$, $p_{2,x}$, $p_{1,y}$, and $p_{2,y}$. The dimensions of the contact hole itself are fixed at:
\begin{align*}
  d_{x}&=800\,\mathrm{nm},\\
  d_{y}&=600\,\mathrm{nm}.
\end{align*}
As incoming light we use conventional illumination \cite{LEV04} which is modelled by a set of incoming plane waves whose incoming angles lie within a cone up to a certain maximum angle. The source is visualized by a set of points in the $k_x$-$k_y$-plane, as shown in Fig. \ref{fig:opcMaskRound}(a). For each of these incoming directions we simulate two orthogonal polarization states to mimic unpolarized light. This gives $P=74$ sources in total. For each of the sources we compute the near field and corresponding far field coefficients separately in order to determine a partially coherent intensity distribution. The far field of the mask passes an optical system with 4 to 1 reduction.

The finite element discretization gives a system with $\nn=474720$ unknowns. The for solution of the problem is about $9,300s$ (single CPU time). Hence optimization of this structure is extremely time using the truth approximation.

\label{sec:examples}
\begin{centering}
\begin{figure}
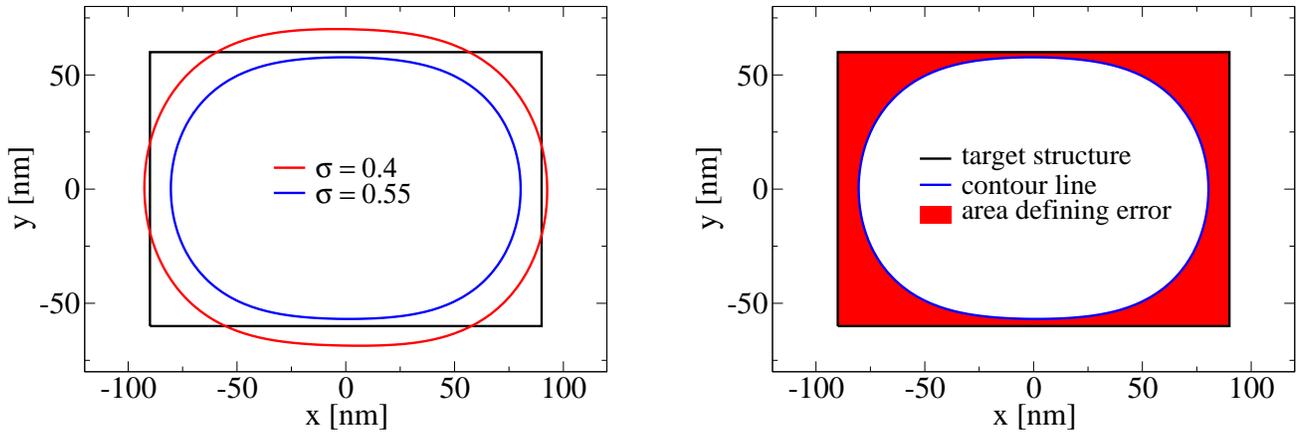

\centering
\psfrag{}{$$}
(a)\hspace{7cm}(b)\\\vspace{0.4cm}
\includegraphics[width=8cm]{fig/opcNoSerifs.eps}\hfill
\includegraphics[width=8cm]{fig/opcErrorFunc.eps}\hfill
\caption{\label{fig:contErrorFunc}
(a) Comparison of contour lines for different levels of $\sigma$. (b) Structure on wafer given as contour line of aerial image at $\sigma=0.55$, i.e. $55\%$ of maximum intensity, with target structure and area defining error functional.
}
\end{figure}
\end{centering}
\begin{centering}
\begin{figure}
\begin{flushleft}
(a)\hspace{8cm}(b)\\\vspace{0.4cm}
\end{flushleft}
\centering
\includegraphics[height=7cm]{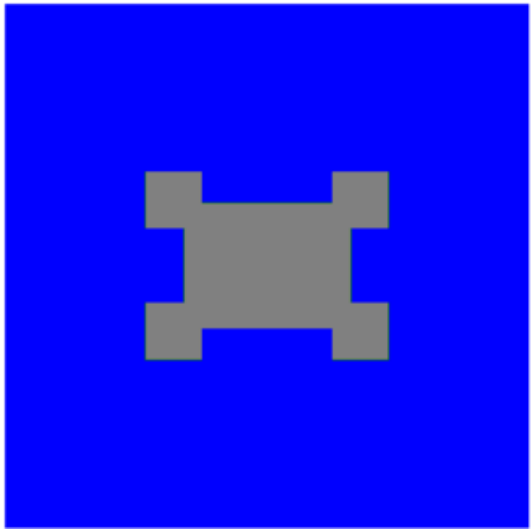}\hfill
\includegraphics[height=7cm]{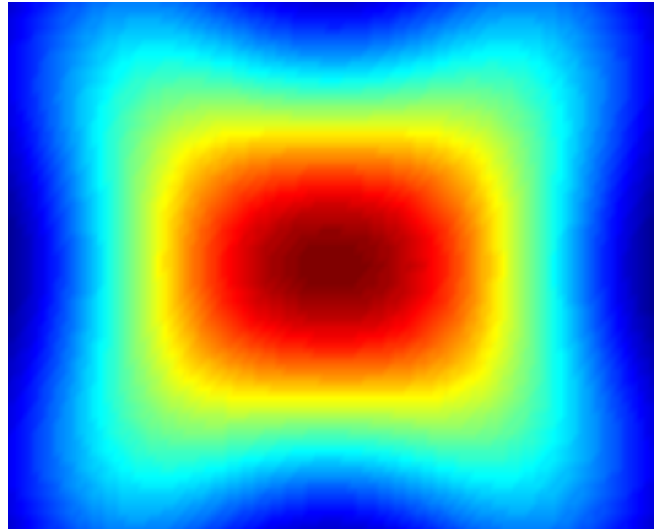}
\caption{\label{fig:opcOptStructure}
(a) Mask layout after OPC optimization and (b) corresponding aerial image (red: high intensity, blue: low intensity).
}
\end{figure}
\end{centering}

\begin{centering}
\begin{figure}
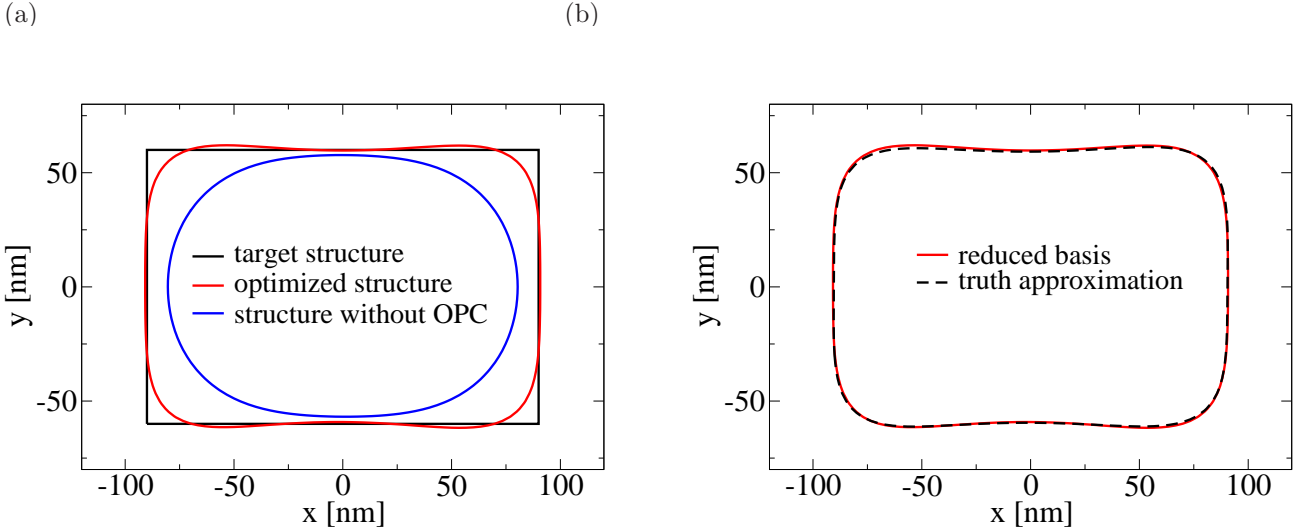

\begin{flushleft}
(a)\hspace{7cm}(b)\\\vspace{0.6cm}
\end{flushleft}
\centering
\includegraphics[width=8cm]{fig/contCompare.eps}\hfill
\includegraphics[width=8cm]{fig/contCompareFEM.eps}\hfill
\caption{\label{fig:contCompare}
(a) Comparison of structure on wafer without and with optimized mask layout obtained from reduced basis computation. (b) Comparison of optimized reduced basis structure and corresponding structure obtained from truth approximation.
}
\end{figure}
\end{centering}

Therefore we construct a reduced basis approximation of above problem. For the parameter domain $D$ we choose
\begin{align}
\begin{split}
  p_{1x}        \in&\;  [145\,\mathrm{nm}; 315\,\mathrm{nm}],\\
  p_{2x}        \in&\;  [100\,\mathrm{nm}; 200\,\mathrm{nm}],\\
  p_{1y}        \in&\;  [45\,\mathrm{nm}; 205\,\mathrm{nm}],\\
  p_{2y}        \in&\;  [100\,\mathrm{nm}; 210\,\mathrm{nm}].\\
\end{split}
\end{align}
The dimension of the constructed reduced basis is $\nrb=40$. 

Despite the large number of sources and outputs of interest the reduced basis computation only takes $1.1s$. This gives a speed up factor of about 8000 compared to the truth approximation. Hence we can perform the online part of the optimization at very low computational costs.

In order to perform the optimization we have to define a cost functional which is minimized. Let us denote by $\gamma(\nu)$ the shape of the contact hole on the wafer at a certain threshold intensity $\sigma$ . Here we use a contour line of the aerial image for simplicity, see Fig.  \ref{fig:contErrorFunc}(a). We fix the threshold at $\sigma=0.55$. Now we define $\Gamma(\nu)$ by the area enclosed by the target shape $\gamma_{T}$ (the rectangle) and $\gamma_{\sigma}(\nu)$, as depicted in Fig. \ref{fig:contErrorFunc}(b). The cost functional is then given by:
\begin{align}
\label{eq:OPCCost}
  g(\nu)&=\norm{1}{L^{1}(\Gamma(\nu))},
\end{align}
and we want to determine optimal parameters such that:
\begin{align}
  \nu_{\mathrm{min}}&=\min_{\nu\in D}g(\nu).
\end{align}
Figure \ref{fig:opcOptStructure} shows the optimized geometry of the photomask and the corresponding aerial image.

The shape of the structure on the wafer without and with optimized serifs is depicted in Fig. \ref{fig:contCompare}(a). The optimized structure shows good agreement to the target structure. Of course corner rounding can not be avoided completely. Furthermore a comparison of the optimal structure computed with the reduced model and obtained from the truth approximation is given in Fig. \ref{fig:contCompare}(b). We observe very good agreement. Largest deviations are of the order of $1\,$nm below \% of the hole CDs. These deviations will further decrease with higher reduced basis dimension.

\section{Conclusions}
\label{sec::conclusions}
We use the reduced basis method for fast solution of geometrically parametrized electromagnetic scattering problems. The reduced basis method is used to construct an approximative reduced system to a parameter dependent finite element problem. Error estimators thereby assure the reliability of the reduced basis solution. 

A challenging numerical optimization example from optical proximity correction demonstrates the performance of the reduced basis method. A speed-up factor of about 8000 compared to the finite element simulation was obtained, enabling a very fast optimization of the OPC structures of a contact hole. The obtained reduced basis result agrees very well with the exact finite element simulation.

\bibliography{/home/numerik/bzfpompl/myBib}
\bibliographystyle{spiebib}

\end{document}